\documentclass[10pt]{article}
\usepackage{caption}
\usepackage[mathscr]{euscript} 
\usepackage{amsthm}
\usepackage{framed,multirow}

\usepackage{latexsym}
\usepackage{adjustbox}
\usepackage{multicol}
\usepackage{soul}
\usepackage{blindtext}
\usepackage{multirow}
\makeatletter

\usepackage{xcolor}

\usepackage{PRIMEarxiv}

\usepackage[utf8]{inputenc} 
\usepackage[T1]{fontenc}    
\usepackage[hidelinks]{hyperref}    
\usepackage{url}         
\usepackage{booktabs}    
\usepackage{amsfonts}    
\usepackage{nicefrac}    
\usepackage{microtype}      
\usepackage{lipsum}
\usepackage{fancyhdr}    
\usepackage{graphicx}    
\graphicspath{{media/}}     

\usepackage{float}
\usepackage{amsmath,amssymb}
\usepackage[ruled,vlined]{algorithm2e}
\usepackage{tikz}
\usetikzlibrary{shapes.geometric, arrows}

\usepackage{grffile}

\usepackage{bbm}
\usepackage{xcolor}

\usepackage{subcaption}
\usepackage[]{graphicx} 

\title{\texttt{CMiNet}: R package for learning the Consensus Microbiome Network}
\usepackage{caption}
\usepackage{subcaption}
\usepackage{fancyhdr}
\usepackage{authblk} 

\author[1]{Rosa Aghdam}
\author[1,2]{Claudia Sol\'{i}s-Lemus\thanks{Corresponding author: \texttt{solislemus@wisc.edu}}}

\affil[1]{Wisconsin Institute for Discovery, University of Wisconsin-Madison, Madison, WI}
\affil[2]{Department of Plant Pathology, University of Wisconsin-Madison, Madison, WI}
\begin{document}

\maketitle
\section*{ABSTRACT}
\textbf{Summary}: Understanding complex interactions within microbiomes is essential for exploring their roles in health and disease. However, constructing reliable microbiome networks often poses a challenge due to variations in the output of different network inference algorithms. To address this issue, we present \texttt{CMiNet}, an R package designed to generate a consensus microbiome network by integrating results from multiple established network construction methods. \texttt{CMiNet} incorporates nine widely used algorithms, including Pearson, Spearman, Biweight Midcorrelation (Bicor), \texttt{SparCC}, \texttt{SpiecEasi}, \texttt{SPRING}, \texttt{GCoDA}, and \texttt{CCLasso}, along with a novel algorithm based on conditional mutual information (\texttt{CMIMN}). By combining the strengths of these algorithms, \texttt{CMiNet} generates a single, weighted consensus network that provides a more stable and comprehensive representation of microbial interactions. The package includes customizable functions for network construction, visualization, and analysis, allowing users to explore network structures at different threshold levels and assess connectivity and reliability. \texttt{CMiNet} is designed to handle both quantitative and compositional data, ensuring broad applicability for researchers aiming to understand the intricate relationships within microbiome communities.\\
\textbf{Availability:} Source code is freely available at \url{https://github.com/solislemuslab/CMiNet}.\\
\textbf{Contact:} solislemus@wisc.edu and rosaaghdam@gmail.com
\section{INTRODUCTION}
In microbiome research, constructing reliable interaction networks is essential for understanding the complex relationships between microbial species and their roles in health and disease \cite{faust2012microbial,peschel2021netcomi,aghdam2023human}. However, selecting a single algorithm often leads to networks that may vary significantly, raising concerns about the reliability of the inferred interactions \cite{kajihara2024networks}. The variation in network structures generated by different algorithms stems from the diverse approaches these methods use to handle compositional data and infer relationships. To address these issues, we propose a consensus network approach that integrates results from multiple well-established algorithms, providing a more stable and representative view of microbiome interactions. Our package, \texttt{CMiNet}, includes nine widely recognized methods to infer microbial relationships: Pearson, Spearman, Bicor, SparCC \cite{friedman2012inferring}, SpiecEas\_MB \cite{kurtz2015sparse}, SpiecEasi\_Glasso \cite{kurtz2015sparse}, SPRING \cite{yoon2019microbial}, GCoDA \cite{fang2017gcoda}, and CCLasso \cite{fang2015cclasso}. Additionally, we developed one algorithm based on conditional mutual information to construct the microbiome network, \texttt{CMIMN}. Below, we provide a brief description of each algorithm's main idea and its approach to network construction.
\begin{itemize}
    \item Pearson: This method calculates the linear correlation between pairs of microbial species, assuming a normal distribution of data.
    \item Spearman: Spearman’s rank correlation assesses the monotonic relationship between species by ranking data points. It is robust to outliers and non-linear relationships.
    \item Biweight midcorrelation (Bicor) is a robust correlation measure that down-weights outliers, making it suitable for noisy microbiome data. It is particularly useful for detecting non-linear relationships.
    \item Sparse Correlations for Compositional data (SparCC) calculates sparse correlations between microbial species. It estimates these correlations by transforming the compositional data into log-ratios and then iteratively approximating the correlations between the underlying absolute abundances. This approach helps to mitigate the issues caused by the compositional nature of the data and focuses on identifying sparse, meaningful interactions between species \cite{friedman2012inferring}.
    \item  Sparse InversE Covariance estimation for Ecological Association and Statistical Inference (SpiecEasi) leverages the Meinshausen-Bühlmann neighborhood selection algorithm to construct ecological networks by estimating sparse inverse covariance matrices (SpiecEasi\_MB). The method assumes sparsity within the ecological network, making it highly applicable to high-dimensional data with a limited number of samples. To enhance robustness and ensure stable network inference, it uses the Stability Approach to Regularization Selection (StARS) to determine an optimal penalty parameter, improving the precision and reliability of the inferred network structure \cite{kurtz2015sparse}.
    \item  SpiecEasi also employing Sparse Inverse Covariance estimation, this variant uses the graphical lasso approach to penalize the inverse covariance matrix, creating a more refined and sparse network. Though computationally demanding, SpiecEasi\_glasso is adept at capturing precise network structures in datasets with numerous variables and fewer observations, making it advantageous for complex microbiome data \cite{kurtz2015sparse}.
    \item Semi-Parametric Rank-based Correlation and Partial Correlation Estimation (SPRING) is a statistical method designed to infer microbial association networks from quantitative microbiome data. It uses a semi-parametric, rank-based approach to handle zero-inflated data and constructs networks by estimating partial correlations that indicate direct microbial interactions. This approach allows SPRING to provide stable, sparse networks, even in complex and zero-heavy microbiome datasets \cite{yoon2019microbial}.
    \item Generalized Co-Occurrence Differential Abundance (GCoDA): Adapts generalized linear modeling for compositional data, incorporating pseudo-counts to stabilize network inference and identify differential interactions between species \cite{fang2017gcoda}. 
    \item Conditional Mutual Information for cotructing Microbiome Network (CMIMN): A novel approach that captures conditional dependencies between species, leveraging mutual information to detect non-linear associations within the microbiome \cite{aghdam2024_CMIMN}.
    \item Correlation Inference for Compositional Data through Lasso (CCLasso): A Lasso-based method that infers correlations within compositional data while mitigating bias from compositional constraints. It performs regularization to identify sparse interactions effectively \cite{fang2015cclasso}.
\end{itemize}
By leveraging a consensus of these algorithms, Consensus Microbiome Network  (\texttt{CMiNet}) algorithm offers a more robust microbiome network, reducing the risk of algorithm-dependent biases and enhancing the reliability of inferred microbial interactions. This package provides the flexibility to incorporate different methods, allowing researchers to obtain comprehensive and credible insights into the microbiome’s complex network structure.

\section*{IMPLEMENTATIONS}
The \texttt{CMiNet} package is implemented in R and is designed to facilitate robust and reproducible microbiome network construction. The package integrates multiple network inference algorithms, each with unique strengths, to generate a consensus microbiome network. This approach helps mitigate biases associated with using a single method, providing a more comprehensive representation of microbial interactions.\\
\textbf{Package Structure}: The core functionality of \texttt{CMiNet} includes several functions for constructing and analyzing consensus networks:
\begin{itemize} 
\item \textbf{CMiNet Function}: Constructs a consensus network using multiple algorithms and outputs a weighted network matrix and an edge list. It allows users to include or exclude specific algorithms and customize their parameters. 
\item \textbf{process\_and\_visualize\_network Function}: Processes the weighted network and visualizes it across different thresholds. This function helps users explore network connectivity at various levels of edge confirmation. 
\item \textbf{plot\_hamming\_distances Function}: Calculates and visualizes Hamming distances between different network matrices. This provides insights into structural differences between networks generated by different algorithms. 
\item \textbf{plot\_network Function}: Generates a plot of the final consensus network based on a user-defined score threshold, displaying only the most significant edges for detailed analysis. 
\end{itemize}

\textbf{Integration of Algorithms:} 
Table \ref{Tab:1} summarizes the algorithms included in the \texttt{CMiNet} package, indicating each method’s compatibility with both quantitative and non-quantitative data (denoted by 'Yes' in the Q column) and outlining the default parameters defined within these algorithms. Users are provided with the flexibility to modify these parameters based on their preferences, allowing them to achieve a sparser network or explore other options. This diverse selection of algorithms and customizable configurations highlights the variability in network outputs, underscoring the importance of a consensus approach for more reliable microbiome network inference.
\begin{table}
\small 
\setlength{\tabcolsep}{4pt} 
\renewcommand{\arraystretch}{1.1} 
\centering
\begin{tabular}{l|c|l}
\hline
\textbf{Algorithm} & \textbf{Q} & \textbf{Parameter} \\
\hline
Pearson           & No           & - \\
Spearman          & No           & - \\
Bicor             & No           & - \\
SparCC            & Yes          & imax = 20, kmax = 10, alpha = 0.1, Vmin = 1e-4 \\
SE\_mb            & No           & method = 'mb', lambda.min.ratio = 1e-2, nlambda = 15, pulsar.params = list(rep.num = 20, ncores = 4) \\
SE\_glasso        & No           & method = 'glasso', lambda.min.ratio = 1e-2, nlambda = 15, pulsar.params = list(rep.num = 50) \\
SPRING            & Yes          & Rmethod = "original", quantitative = TRUE, ncores = 5, lambdaseq = "data-specific", nlambda = 15, rep.num = 20 \\
GCoDA             & Yes          & counts = FALSE, pseudo = 0.5, lambda.min.ratio = 1e-4, nlambda = 15, ebic.gamma = 0.5 \\
CMIMN             & Yes          & quantitative = TRUE, q1 = 0.7, q2 = 0.95 \\
CCLasso           & Yes          & counts = FALSE, pseudo = 0.5, k\_cv = 3, lam\_int = c(1e-4, 1), k\_max = 20, n\_boot = 20 \\
\hline
\end{tabular}
\caption{Selected algorithms in the \texttt{CMiNet} package, highlighting their suitability for analyzing both quantitative and non-quantitative data (column Q) and detailing relevant parameters.}
\label{Tab:1}
\end{table} 

The \texttt{CMiNet} package covers both correlation-based methods (e.g., Pearson, Spearman, Bicor) and more complex compositional data-specific methods (e.g., SparCC, SpiecEasi, CCLasso). The package also includes a novel conditional mutual information-based algorithm (\texttt{CMIMN}) for detecting non-linear relationships.

\textbf{Parameter Customization} Users can adjust algorithm-specific parameters to suit their data and research needs. For instance, parameters like \texttt{lambda.min.ratio} and \texttt{nlambda} for the SpiecEasi algorithm or \texttt{q1} and \texttt{q2} for \texttt{CMIMN} can be modified directly in the function call. This customization supports a detailed exploration of network structures.

\textbf{Output and Visualization} The package outputs include: \begin{itemize} 
\item A weighted consensus network matrix. 
\item An edge list with edge weights for detailed network analysis. 
\item Visualization plots that illustrate network structures based on selected thresholds and highlight key nodes and connections. \end{itemize}
These outputs help researchers interpret microbiome interactions at various levels of stringency, providing a flexible platform for examining microbial relationships.

\begin{figure}
    \centering
    \begin{subfigure}{0.48\textwidth}
        \includegraphics[width=\linewidth]{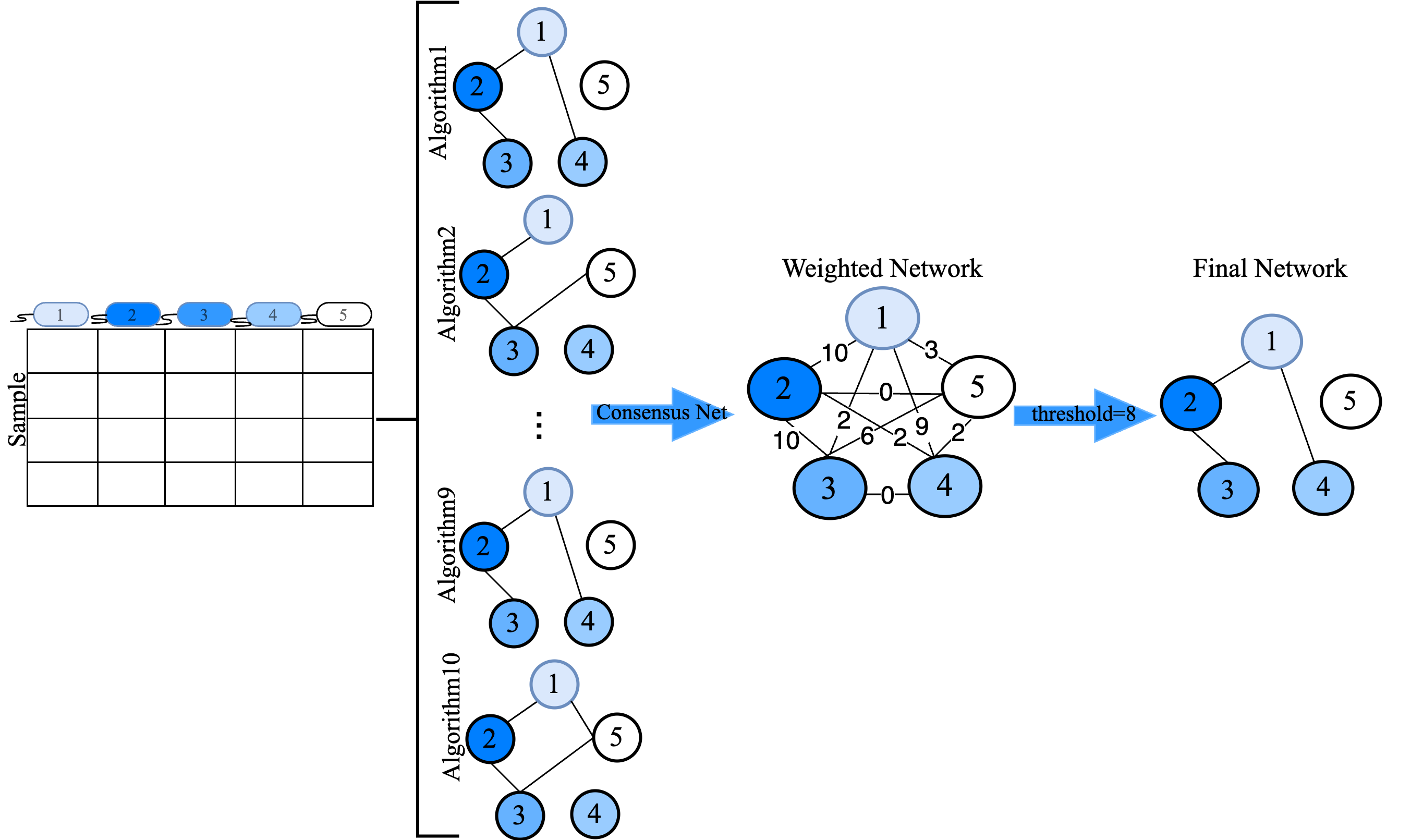}
        \caption{Overview of \texttt{CMiNet}}
        \label{fig:part_a}
    \end{subfigure}
    \hfill
    \begin{subfigure}{0.45\textwidth}
        \includegraphics[width=\linewidth]{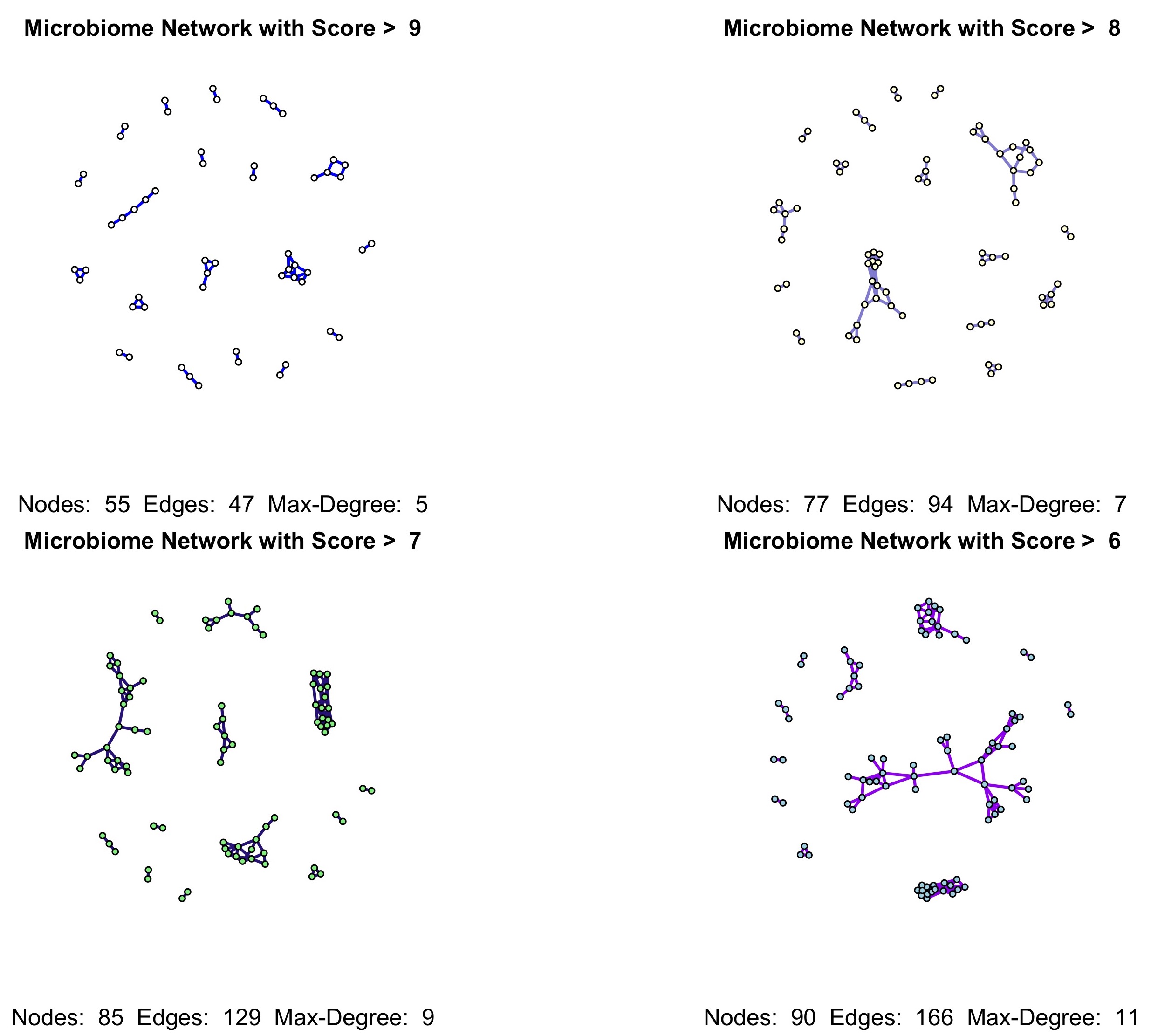}
        \caption{Results of applying \texttt{CMiNet} }
        \label{fig:part_b}
    \end{subfigure}
    \caption{Illustration of the \texttt{CMiNet} package and its results. (a) presents the concept of the \texttt{CMiNet} package, demonstrating how it integrates multiple algorithms to construct a consensus microbiome network. (b) shows the results of applying \texttt{CMiNet} to microbiome data, depicting networks generated at varying threshold levels. For example, with a threshold of 9 (where only edges confirmed by all ten methods are included), the resulting network comprises 55 nodes and 47 edges. Lowering the threshold to 8 (including edges confirmed by nine or all ten methods) results in a network with 77 nodes and 94 edges.}
    \label{fig:combined}
\end{figure}

\textbf{Example Application:} 
Figure \ref{fig:combined} provides an overview of the \texttt{CMiNet} package and its results. Figure (\ref{fig:part_a}) illustrates the concept behind \texttt{CMiNet} for learning a consensus microbiome network by integrating multiple algorithms. Figure (\ref{fig:part_b}) displays the results of running \texttt{CMiNet} on microbiome data, showing networks generated at different threshold levels. For example, using a threshold value of 9, which includes only edges confirmed by all ten methods, results in a network with 55 nodes and 47 edges. Lowering the threshold to 8, which includes edges confirmed by at least nine out of ten methods, expands the network to 77 nodes and 94 edges. By considering this figure, users can set a threshold based on the number of nodes and edges and construct a final network that includes only edges with weights higher than the defined threshold. This flexibility allows users to customize their analyses to match the level of confidence they wish to achieve. This approach highlights the importance of considering edges confirmed by all or most algorithms rather than relying on a single method. Given the significant differences in the outputs of various algorithms and the lack of a gold standard in microbiome research for evaluating relationships between microbes, using a consensus approach ensures more robust and reliable network inference.

\section*{Conclusion} 
Understanding microbiome networks is crucial for studying the complex interactions within microbial communities and their implications for health and disease\cite{nelson2024minaa}. One of the significant challenges in this field is the absence of a gold standard for evaluating the accuracy of network inference algorithms. This lack of a benchmark makes it difficult to assess the reliability of networks produced by different methods, as various algorithms often yield substantially different results. Additionally, certain algorithms may only support specific data types, such as quantitative or normally distributed data, which further complicates the selection of an appropriate method for a given dataset.

The \texttt{CMiNet} package addresses these challenges by integrating multiple well-established algorithms to construct a consensus microbiome network. By leveraging diverse methods—ranging from simple correlation-based approaches to advanced techniques that account for conditional dependencies—\texttt{CMiNet} offers a comprehensive framework that enhances the reliability of inferred microbial interactions. This integration ensures that the network reflects a more balanced view, mitigating biases that can result from relying on a single algorithm.

The importance of using multiple methods lies in their varied assumptions and strengths; combining correlation-based methods with those that capture conditional dependencies provides a richer, more accurate depiction of microbial relationships. The consensus approach facilitates robust analyses, allowing researchers to compare results against a more consistent and representative network. The outputs of \texttt{CMiNet} provide a reliable foundation for subsequent studies and can serve as a benchmark for comparing newly developed algorithms.

Looking ahead, we plan to expand the functionality of \texttt{CMiNet} by developing an R Shiny application. This app will enable users to easily upload their data, select algorithms, and generate consensus networks and visualizations with minimal technical expertise. This future enhancement aims to make robust microbiome network construction more accessible, supporting broader applications in microbiome research and facilitating discoveries in microbial ecology and personalized medicine.

\section*{Acknowledgements}
Thank you to Dr. Maryam Shahdoost for checking, installing, and running the \texttt{CMiNet} package.

\section*{Conflicts of interest}
None declared.
\section*{Funding}
This work was supported by the National Science Foundation (DEB-2144367 to CSL).

\bibliographystyle{unsrt}
\bibliography{ref}
\end{document}